\documentclass[final,conference]{IEEEtran}
\IEEEoverridecommandlockouts
\usepackage[utf8]{inputenc} 
\usepackage[T1]{fontenc}
\usepackage{url}
\usepackage{ifthen}
\usepackage[cmex10]{amsmath}
\usepackage{array} % Use the [cmex10] option to ensure complicance
\usepackage{enumerate}
\usepackage{amssymb}
\usepackage{amsmath} 
\usepackage[lined,boxed,commentsnumbered, ruled]{algorithm2e}
\usepackage{mathrsfs}
\usepackage{bm}
\usepackage{tikz}
\usepackage{algorithmic}
\usetikzlibrary{arrows}
\usepackage{subfigure}
\usepackage{graphicx,booktabs,multirow}
\usepackage{epstopdf}
\usepackage{subfigure}
\usepackage{multirow}
\usepackage{tabularx} 
\usepackage{booktabs}

\definecolor{colorhkust}{RGB}{20,43,140}
\definecolor{colortsinghua}{RGB}{116,52,129}
\definecolor{color1}{RGB}{128,0,0}

\usepackage{amsthm}
\usepackage{verbatim}
\usepackage{color}

\newtheorem{proposition}{Proposition}

\newcommand{\diagg}{\mathrm{diag}}

\begin{document}

        \title{Reconfigurable Intelligent Surface Enhanced Cognitive Radio Networks}
        
\author{Jinglian He$^{*\dag\S}$, Kaiqiang Yu$^*$, Yong Zhou$^*$, and Yuanming Shi$^*$, \textit{Member}, \textit{IEEE},\\ 
        
        \IEEEauthorblockA{
                $^{*}$School of Information Science and Technology,
                ShanghaiTech University, Shanghai 201210, China\\
                $^{\dag}$Shanghai Institute of Microsystem and Information Technology, Chinese Academy of Sciences, China\\
                $^{\S}$University of Chinese Academy of Sciences, Beijing 100049, China\\
                E-mail: \{hejl1, zhouyong, shiym\}@shanghaitech.edu.cn, yukaiqiangsdu@gmail.com}\\
}
        \maketitle
\begin{abstract}
        The cognitive radio (CR) network is a promising network architecture that meets the requirement of enhancing scarce radio spectrum utilization. Meanwhile, reconfigurable intelligent surfaces (RIS) is a promising solution to enhance the energy and spectrum efficiency of wireless networks by properly altering the signal propagation via tuning a large number of passive reflecting units. In this paper, we investigate the downlink transmit power minimization problem for the RIS-enhanced single-cell cognitive radio (CR) network coexisting with a single-cell primary radio (PR) network by jointly optimizing the transmit beamformers at the secondary user (SU) transmitter and the phase shift matrix at the RIS. The investigated problem is a highly intractable due to the coupled optimization variables and unit modulus constraint, for which an alternative
minimization framework is presented. Furthermore, a novel difference-of-convex (DC) algorithm is developed to solve the resulting non-convex quadratic program by lifting it into a low-rank matrix optimization problem. We then represent non-convex rank-one constraint as a DC function by exploiting the difference between trace norm and spectral norm. The simulation results validate that our proposed algorithm outperforms the existing state-of-the-art methods.
        
\end{abstract}

\section{Introduction}
The radio spectrum is a scarce natural resource in wireless communication systems \cite{ali2016advances}. The great majority of available radio spectrum has been licensed by governments \cite{xu2020resource}. To provide high data-rate communication services for the explosive growth of mobile data traffic in 6G network \cite{6g}, the spectrum scarcity problem can't wait to be solved. The CR network holds the reliable and effective promise of meeting the spectrum scarcity problem \cite{mitola1999cognitive}. In CR network, the unlicensed SUs are allowed to operate within the service range of the primary users (PUs) as long as the PUs can be well protected \cite{gharavol2010robust}. A great deal of literatures \cite{ali2016advances, mitola1999cognitive,gharavol2010robust} demonstrated the good performance of CR network. Although CR network brings a quantum leap in spectrum efficiency of wireless networks,  the network energy consumption and hardware cost are still serious challenging in practical applications \cite{zhang2016fundamental}. To this end, realizing green and sustainable \cite{shi2014group} next generation wireless network need focus on finding both spectral and energy efficient techniques with low hardware cost \cite{wu2017overview}.

 Reconfigurable intelligent surfaces (RIS) is an attractive technology to provide green and cost-effective solution for substantially enhancing the energy and spectrum efficiency of wireless networks. RIS is a planar array consisting of a large number of passive elements, each of which is able to induce a certain phase shift (by a smart controller) independently on the incident signal without employing any power amplifier \cite{huang2019reconfigurable}. Thus there is no additional thermal noise added during reflecting and RIS consumes extremely low energy.
 As a result, RIS-enhanced CR network is considered to be applied for improving spectral efficiency and energy efficiency \cite{yuan2019intelligent}. However, the designs for other RIS works which can solved by manifold optimization are not suitable for RIS-enhanced CR network \cite{xu2020resource}. Since the RIS-enhanced CR network are more complex and the RIS-enhanced CR network optimization problems are more intractable. Motivated by this challenge, this paper focuses on the open issue, which requires the joint design of beamforming vectors at the SU and phase shift matrix at RIS  for an RIS-enhanced CR network. 

This paper considers a downlink RIS-enhanced CR network. We propose to jointly optimize the beamforming vectors at the SU and the phase shift matrix at the RIS to minimize the total transmit power consumption at the SU, while satisfying the SU interference power constraints at each PU and the signal-to-interference-plus-noise ratio (SINR) constraints of each user. To decouple the optimizing variables, an alternative optimization framework is presented, yielding two optimization subproblems. The first subproblem of optimizing beamforming vectors at the SU is a second order cone program (SOCP) and can be efficiently solved. As for the other quadratically constrained quadratic programming (QCQP) subproblem of optimizing the phase shift matrix at the RIS, it can be first lifted into rank-one constrained matrix optimization problem. Conventional approach \cite{ma2010semidefinite} drops the resulting rank-one constraints to obtain a semidefinite programming (SDP) problem and uses the semidefinite relaxation (SDR) technique, yielding poor performance solutions in the high-dimensional settings \cite{yang2018federated, jiang2019over}. To address the challenge of SDR, the resulting rank-one constraint is rewritten by exact DC representation in this paper. An efficient DC algorithm is presented to deal with the resulting non-convex DC programming problem. The simulation results demonstrate the good performance of the proposed DC approach compared with the existing methods in the RIS-enhanced CR network.

\section{System Model and Problem Formulation}
\subsection{System Model}

\begin{figure}[t]
        \centering
        \includegraphics[scale = 0.30]{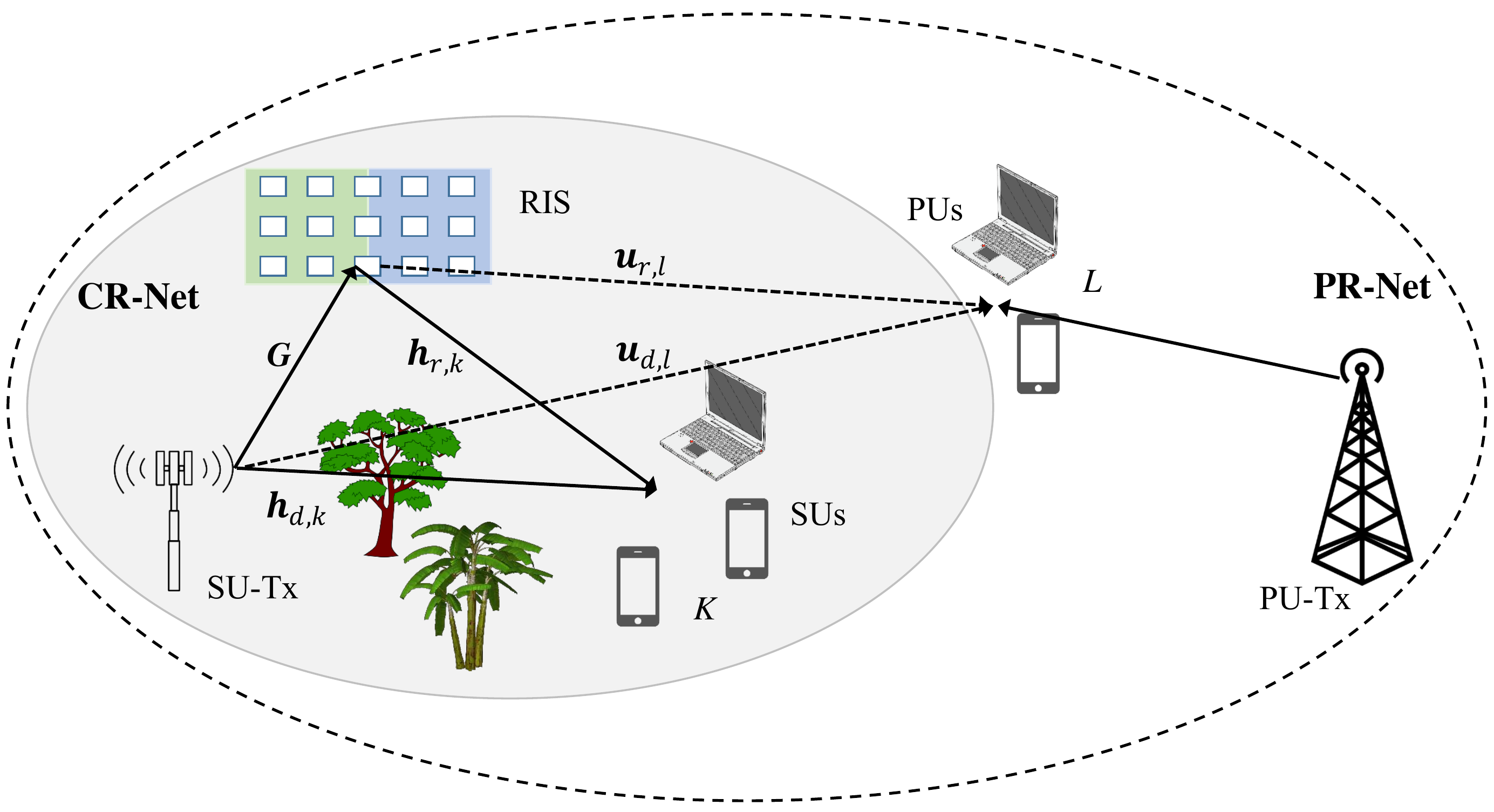}
        \vspace{-0.10in}
        \caption{An RIS-enhanced single-cell CR network coexisting with a single-cell PR network.}
        \label{model}
        \vspace{-0.20in}
\end{figure}

In this paper, we consider the downlink transmission of a multi-user MISO cognitive ratio network consisting of a $N$-antenna SU transmitter and $K$ single-antenna users shown in Fig. \ref{model}, where an RIS with $M$ passive reflecting elements is equipped to enhance the transmission. Due to higher efficiency of spectrum utility and interference management, we thus adopt the spectrum sharing operation model for proposed CR networks. Specifically, the SUs are allowed to transmit concurrently with $L$ single-antenna PUs in the coexisting primary ratio network at the same band. Moreover, the RIS can smartly change phase shifts by operating in two modes, i.e., the receiving mode for estimating channel state information (CSI) and the reflecting mode for scattering incident signals. For the ease of system design, we further focus on a scenario where the PU transmitter is deployed far enough from proposed CR network. Therefore, the interfering power received from PU transmitter can be regarded as noises \cite{arxiv2020resource}. The system thus consists of five sets of channels, i.e., transmitter-RIS link, transmitter-SU link, transmitter-PU link, RIS-SU link and RIS-PU link, for which we further assume a quasi-static flat-fading channel model with prefect CSI. 

Let $s_k\in\mathbb{C}$ and $\bm{w}_k\in\mathbb{C}^N$ denote the transmitted symbol and linear beamforming vector at SU transmitter for $k$-th SU. It is assumed that $s_k$ is zero mean with $\mathbb{E}[|s_k|^2]=1$. The received signal at $k$-th SU is given by \cite{gharavol2010robust,arxiv2020resource}
\begin{eqnarray}
        %y_k=(\bm{h}_{d,k}^{\sf{H}}+\bm{h}_{r,k}^{\sf{H}}\bm{\Theta}\bm{G})\sum_{i=1}^{K}\bm{w}_is_i+n_k,\ \forall k\in\mathcal{K},
        y_k\!\!=\!\!(\bm{h}_{d,k}^{\sf{H}}\!+\!\bm{h}_{r,k}^{\sf{H}}\bm{\Theta}\bm{G})\bm{w}_ks_k\!\!+\!\!(\bm{h}_{d,k}^{\sf{H}}\!+\!\bm{h}_{r,k}^{\sf{H}}\bm{\Theta}\bm{G})\!\sum_{i\neq k}\!\!\bm{w}_is_i\!+\!n_k,\!\!\!\!\!
\end{eqnarray}
where $k\in\mathcal{K}=\{1,2,...,K\}$, $n_k$ is the additive white Gaussian noise with distribution $\mathcal{CN}(0,\sigma^2)$, and $\bm{h}_{d,k}\in\mathbb{C}^N$, $\bm{h}_{r,k}\in\mathbb{C}^M$ and $\bm{G}\in\mathbb{C}^{M\times N}$ respectively denote the channel from the SU-Tx to $k$-th SU, from the RIS to $k$-th SU and from the SU-Tx to the RIS. Besides, let $\theta_m \in \mathbb{C}$ denote the reflection coefficient of the $m$-th RIS element. We assume that $|\theta_m|=1$ and the phase of $\theta_m$ can be flexibly adjusted in $[0,2\pi)$.  The matrix $\bm{\Theta}=\beta\diagg({\theta_1,...,\theta_M)}$ represents the diagonal phase shifts matrix of the RIS. Without loss of generality, we further assume $\beta=1$. The SINR at $k$-th SU can be written as
\begin{eqnarray}
\text{SINR}_k=\frac{|(\bm{h}_{d,k}^{\sf{H}}+\bm{h}_{r,k}^{\sf{H}}\bm{\Theta}\bm{G})\bm{w}_k|^2}{\sum_{i\neq k}|(\bm{h}_{d,i}^{\sf{H}}+\bm{h}_{r,i}^{\sf{H}}\bm{\Theta}\bm{G})\bm{w}_i|^2+\sigma^2}.
\end{eqnarray}
In addition, let $\bm{u}_{d,l}\in\mathbb{C}^N$ and $\bm{u}_{r,l}\in\mathbb{C}^M$ denote the channel from the SU-Tx to $l$-th PU and from the RIS to $l$-th PU, respectively. The received signal at $l$-th PU is given by \cite{gharavol2010robust}
\begin{eqnarray}
                z_l=s_l^{P}+(\bm{u}_{d,l}^{\sf{H}}+\bm{u}_{r,l}^{\sf{H}}\bm{\Theta}\bm{G})\sum_{i=1}^{K}\bm{w}_is_i+n_l,\ \forall l\in\mathcal{L},
\end{eqnarray} 
where $\mathcal{L}=\{1,2,...,L\}$. We further give the SU interference power at $l$-th PU as \cite{gharavol2010robust}
\begin{eqnarray}
        \text{IP}_l=\sum_{k=1}^K|\bm{u}_l^{\sf{H}}\bm{w}_k|^2,\ \forall l\in\mathcal{L},
\end{eqnarray}
where $\bm{u}_{l}^{\sf{H}}=\bm{u}_{d,l}^{\sf{H}}+\bm{u}_{r,l}^{\sf{H}}\bm{\Theta}\bm{G}$.

\subsection{Problem Formulation}
  In this paper, we aim to minimize the transmit power via jointly optimizing the beamforming vectors $\bm{w}_k$ at the SU-Tx and the phase shift matrix $\bm{\Theta}$ at the RIS, while guaranteeing the QoS requirements of each SUs. Furthermore, we need manage the interference induced by the simultaneous transmission in proposed CR network to protect PUs. Following one of piratical design paradigms of decentralized approach, we propose to impose a constraint on the maximum SU interference power at PUs. Then, the transmitted power minimization problem can be reformulated as
  \setlength\arraycolsep{2pt}
  \begin{eqnarray}\label{P0}
  \mathscr{P}:\ \mathop{\text{minimize}}_{\{\bm{w}_k\},\bm{\Theta}} && \sum_{k=1}^K\|\bm{w}_k\|^2 \nonumber\\
  \textrm{subject to} && \text{SINR}_k\geq \gamma_k,\ \forall k\in\mathcal{K},\\
  &&\text{IP}_l\leq \kappa_l,\ \forall l\in\mathcal{L},\\
  &&|\theta_m| =1 , \forall \, m = 1,\ldots, M,
  \end{eqnarray}
  where $\|\bm{w}_k\|^2$ denotes the transmit power for $k$-th SU, $\gamma_k$ represents the minimum required SINR for $k$-th SU and $\kappa_l$ is the upper limit of SU interference power at $l$-th PU.
  
  Unfortunately, this problem turns out to be highly intractable due to the unit modulus constraint as well as the coupled phase shift matrix $\bm{\Theta}$ and beamforming vectors $\bm{w}_k$. To solve this problem, we then present an alternating optimization approach in next section.
  
  \section{Alternating Optimization Algorithm}
  In this section, an alternating optimization framework is presented, where beamforming vectors $\bm{w}_k$ and phase shift matrix $\bm{\Theta}$ are separately and iteratively optimized with another fixed, until the convergence is achieved.
  
  \subsection{Optimizing Beamforming Vectors}
  With a given phase shift matrix $\bm{\Theta}$, channel responses $\bm{h}_{k}^{\sf{H}}=\bm{h}_{d,k}^{\sf{H}}+\bm{h}_{r,k}^{\sf{H}}\bm{\Theta}\bm{G}$ and $\bm{u}_{l}^{\sf{H}}$ are fixed, and thus problem $\mathscr{P}$ can be written as the following non-convex problem
  \begin{eqnarray}\label{w}
\mathop{\text{minimize}}_{\{\bm{w}_k\}} && \sum_{k=1}^K\|\bm{w}_k\|^2 \nonumber\\
  \textrm{subject to} &&\label{sinr} \text{SINR}_k\geq \gamma_k,\ \forall k\in\mathcal{K},\nonumber\\
  &&\text{IP}_l\leq \kappa_l,\ \forall l\in\mathcal{L}.
  \end{eqnarray}
  It has been showed that problem (\ref{w}) is a power minimization problem in conventional CR downlink network, which is an SOCP  problem and can be solved efficiently by using interior point methods \cite{Tajer2009Beamforming,Boyd2004Convex}.

  \subsection{Optimizing Phase Shift Matrix}
  With given beamforming vectors $\{\bm{w}_k,\forall k \in \mathcal{K}\}$, let $b_{k} = \bm h_{d,k}^{\sf H}\bm w_k$, $\forall k\in \mathcal{K}$, and $\bm \theta^{\sf H}\bm a_{k} = \bm h_{r,k}^{\sf H}\bm \Theta{\bm G}\bm w_k$, where $ \bm \theta=[\theta_1,\ldots,\theta_M] ^{\sf H}$ and $\bm a_{k} = {\diagg}(\bm h_{r,k}^{\sf H})\bm G\bm w_k $. Let $c_{l,k} = \bm u_{d,l}^{\sf H}\bm w_k$, and $\bm \theta^{\sf H}\bm d_{l,k} = \bm u_{r,l}^{\sf H}\bm \Theta{\bm G}\bm w_k$, $\bm d_{l,k} = {\diagg}(\bm u_{r,l}^{\sf H})\bm G\bm w_k$,$\forall l\in\mathcal{L}$, $\forall k\in \mathcal{K}$. Consequently, we can rewritten problem $\mathscr{P}$ as the following nonconvex feasibility detection problem 
  \begin{eqnarray}\label{theta}
 \mathop{\text{Find}}
  &&{\bm \theta}\nonumber \\
\textrm{subject to}&& \gamma_k \big(\!\sum_{i\neq k}^{K}|\bm \theta^{\sf H} \bm a_{i}\!+\!b_{i}|^2\!+\!\sigma^2\big)\!\leq|\bm \theta^{\sf H} \bm a_{k}\!+\!b_{k}|^2,  \forall \, k\in \mathcal{K},\nonumber\\
  &&\sum_{k=1}^{K}|c_{l,k}+\bm{\theta}^{\sf H}\bm{d}_{l,k}|^2\leq \kappa_l,\forall l\in\mathcal{L},\nonumber\\
  &&|\theta_m| =1 , \forall \, m = 1,\ldots, M.
  \end{eqnarray}
  Despite that problem (\ref{theta}) is non-convex and inhomogeneous, we can introduce an auxiliary variable $t$ and reformulate problem (\ref{theta}) to be a homogenous non-convex QCQP problem. Hence, the reformulated problem is given by
   \begin{eqnarray}\label{theta1}
 \mathop{\text{Find}}
 &&{\bm{\tilde{\theta}}}\nonumber \\
\textrm{subject to}&&\gamma_k \big(\!\sum_{i\neq k}^{K}\bm{\tilde{\theta}}^{\sf H} \bm R_{i}\bm{\tilde{\theta}}\!+\!b_{i}^2\!+\!\sigma^2\big) \nonumber\!\leq\!\bm {\tilde{\theta}}^{\sf H} \bm R_{k}\bm{\tilde{\theta}}\!+\!b_{k}^2,  \forall \, k\in \mathcal{K},\nonumber\\
  &&\sum_{k=1}^{K}c_{l,k}^2+\bm{\tilde{\theta}}^{\sf H}\bm{Q}_{l,k}\bm{\tilde{\theta}}\leq \kappa_l,\forall l\in\mathcal{L},\nonumber\\
  &&| \theta_m| =1 , \forall \, m = 1,\ldots, M,
  \end{eqnarray}
  where
  \begin{eqnarray}
  \!\!\!\!\!\!\!\!\!\!\!\bm R_{k}\!\!=\!\!
  \begin{bmatrix}
  \bm a_{k}\bm a_{k}^{\sf H}  &\bm a_{k}  b_{k}      \\
  b_{k}^{\sf H}\bm a_{k}^{\sf H} \!\! &\!\! 0 
  \end{bmatrix}, 
  \bm Q_{l,k}\!\!=\!\!
  \begin{bmatrix}
  \bm d_{l,k}\bm d_{l,k}^{\sf H}&\bm d_{l,k}     c_{l,k}      \\
  c_{l,k}^{\sf H}\bm d_{l,k}^{\sf H} \!\!& 0 
  \end{bmatrix}, {\bm {\tilde{\theta}}}\!\!=\!\!
  \begin{bmatrix}
  \bm {\theta}    \\
  t
  \end{bmatrix}.
  \end{eqnarray}
  We denote a feasible solution to problem (\ref{theta1}) as $\bm{\tilde{\theta}}^*$.  Let $\bm{\theta}=[\bm{\tilde{\theta}}^*/\bm{\tilde{\theta}}_{M+1}^*]_{1:M}$ be a feasible solution to problem (\ref{theta}), where $[\bm{z}]_{1:M}$ is the first M elements of $\bm{z}$.
 
We exploit the matrix lifting technique to cope with the non-convex quadratic constraints in problem (\ref{theta1}). Let $\bm{\tilde{\Theta}}=\bm{\tilde{\theta}}\bm{\tilde{\theta}}^{\sf H}$, and $\text{Tr}(\bm{R}_k\bm{\tilde{\Theta}})=\bm {\tilde{\theta}}^{\sf H} \bm R_{k}\bm{\tilde{\theta}}$. Thus, we can rewrite problem (\ref{theta1}), yielding rank-one constrained matrix feasibility problem:
 \begin{eqnarray}\label{Theta1}
 \mathop{\text{Find}}
&&{\bm{\tilde{\Theta}}}\nonumber \\
\textrm{subject to}\!\!&&\!\gamma_k \big(\!\sum_{i\neq k}^{K}\!\!\text{Tr}(\!\bm{R}_i\bm{\tilde{\Theta}}\!)\!+\!b_{i}^2\!+\!\sigma^2\big)\!\leq\!\text{Tr}(\!\bm{R}_k\bm{\tilde{\Theta}}\!)\!+\!b_{k}^2, \forall \, k\in \mathcal{K},\nonumber\\
&&\sum_{k=1}^{K}c_{l,k}^2+\text{Tr}(\bm{Q}_{l,k}\bm{\tilde{\Theta}})\leq \kappa_l,\forall l\in\mathcal{L},\nonumber\\
&&\bm{\tilde{\Theta}}_{m,m} =1 , \forall \, m = 1,\ldots,M+1,\nonumber\\
&&\bm{\tilde{\Theta}} \succcurlyeq 0,\mathrm{rank}(\bm{\tilde{\Theta}})\!=\!1.
\end{eqnarray}
  
  As a result, problem (\ref{Theta1}) is still non-convex and highly intractable due to the nonconvex rank constraint, i.e., $\mathrm{rank}(\bm{\tilde{\Theta}})=1$. To this end, the SDR technique is naturally used to cope with the non-convex rank constrain \cite{ma2010semidefinite}. Since after dropping the nonconvex rank constraint, the yielding SDP problem can be directly solved by the existing solvers. While the solver returns the solution of the relaxed SDP problem which fails to be the anticipated rank-one solution, we can obtain a suboptimal solution by adopting the Gaussian randomization \cite{ma2010semidefinite}. However, the SDR technique also has its limitation that when the dimension of the optimization parameters is high, the probability of obtaining the anticipated rank-one solution is small \cite{yang2018federated, jiang2019over}.
  
  To deal with the drawback of the SDR technique, we propose an exact DC representation with respect to the rank-one constraint of the optimization variable matrix. The difference between the trace norm and the spectral norm is used to find the DC representation in the following section.
 
  \section{Alternating DC Approach}
 In this section, an exact DC representation is presented for the rank-one constraint by leveraging the difference of two convex norms. Then we propose an efficient alternating DC algorithm to solve the resulting DC programming.
  
  \subsection{DC Representation for Rank Function}
  First of all, a novel DC representation is presented for the original rank-one constraint of problem (\ref{Theta1}) in the following proposition \cite{yang2018federated}.
  \begin{proposition} 
        For PSD matrix $\bm \Psi\in \mathbb{C}^{N\times N}$ and $\mathrm{Tr}(\bm \Psi) >0$, we have
          \vspace{-0.10in}
        \begin{equation*}
        \mathrm{rank}(\bm \Psi)=1\Leftrightarrow \mathrm{Tr}(\bm \Psi)-\|\bm \Psi\|_2=0,
        \end{equation*}
        where  trace norm $\mathrm{Tr}(\bm \Psi)=\sum_{i=1}^{N} \sigma_i(\bm \Psi)$ and spectral norm $\|\bm \Psi\|_2=\sigma_1(\bm \Psi)$ with $ \sigma_i(\bm \Psi) $ denoting the $ i $-th largest singular value of matrix $ \bm \Psi $. 
  \end{proposition}
  It is not difficult to verify that in problem (\ref{Theta1}), the lifting matrix $\bm{\tilde{\Theta}}$ has rank one. Then, we can apply the DC representation to reformulate the original rank-one matrix $\bm{\tilde{\Theta}}$ constraint in problem (\ref{Theta1}) as the difference between the trace norm and the spectral norm, the reformulated problem (\ref{Theta1}) is given by
\begin{eqnarray}\label{DC}
\mathop{\text{minimize}}
  &&\text{Tr}(\bm {\tilde{\Theta}})-\|\bm {\tilde{\Theta}}\|_2\nonumber \\
  \textrm{subject to} &&\gamma_k \big(\sum_{i\neq k}^{K}\text{Tr}(\bm{R}_i\bm{\tilde{\Theta}})+b_{i}^2+\sigma^2\big)\leq\text{Tr}(\bm{R}_k\bm{\tilde{\Theta}})+b_{k}^2, \nonumber\\
  && \forall \, k\in \mathcal{K},\nonumber\\
  &&\sum_{k=1}^{K}c_{l,k}^2+\text{Tr}(\bm{Q}_{l,k}\bm{\tilde{\Theta}})\leq \kappa_l,\forall l\in\mathcal{L},\nonumber\\
  &&\bm{\tilde{\Theta}}_{m,m} =1 , \forall \, m = 1,\ldots,M+1,\nonumber\\
  &&\bm{\tilde{\Theta}} \succcurlyeq 0.
  \end{eqnarray}
  Definitely, when the value of the problem (\ref{DC})’s objective function becomes zero, which means that we obtain an exact rank-one solution. When a rank-one optimal solution $\bm{\tilde{\Theta}}^*$ is obtained, the feasible solution of RIS reflection coefficient vector $\bm{\tilde{\theta}}^*$ to problem (\ref{theta1}) can be attained by Cholesky decomposition $\bm{\tilde{\Theta}}^*=\bm{\tilde{\theta}}^*\bm{\tilde{\theta}}^{*\sf H}$ \cite{Fu2019intelligent}. Additionally, we can directly claim that the original problem (\ref{theta}) is infeasible when the objective value of the function fails to be zero \cite{Fu2019blind}.
  
  \subsection{DC Algorithm for Problem $(\ref{DC})$}
  Despite the fact that the DC programming problem $(\ref{DC})$ is still non-convex due to their objective functions. Hopefully, we can exploit the good structure of the objective function to design efficient algorithm by solving the convex relaxation versions of the primal and dual problems of DC programming. Thus the problem $(\ref{DC})$ can be rewritten as 
    \vspace{-0.10in}
  \begin{eqnarray}\label{DC1}
  \mathop{\text{minimize}}_{\bm{\tilde{\Theta}}}
  &&\text{Tr}(\bm{\tilde{\Theta}})-\|\bm{\tilde{\Theta}}\|_2 + \Delta_{\mathcal{C}}(\bm{\tilde{\Theta}}),
  \end{eqnarray}
  where $\mathcal{C}$ is defined as convex sets that satisfy the constraints in problem $(\ref{DC})$, and the indicator function $\Delta_{\mathcal{C}}(\bm{\tilde{\Theta}})$ is defined as
  $$\Delta_{\mathcal{C}}(\bm{\tilde{\Theta}})=
  \begin{cases}
  0,& \bm{\tilde{\Theta}}\in \mathcal{C}\\
  +\infty,& \text{otherwise}
  \end{cases}.$$
  It is not difficult to verify that problems (\ref{DC1}) have the special structure of minimizing the difference of two convex functions, which is given by
  \begin{eqnarray}\label{DC structure}
  \mathop{\textrm{minimize}}_{\bm{\tilde{\Theta}}} &&f = p(\bm{\tilde{\Theta}}) - q(\bm{\tilde{\Theta}}). 
  \end{eqnarray} 
  In problem (\ref{DC1}), $p(\bm{\tilde{\Theta}})$ denotes the function $\text{Tr}(\bm{\tilde{\Theta}})+\Delta_{\mathcal{C}}(\bm{\tilde{\Theta}})$, and $q(\bm{\tilde{\Theta}})$ denotes the function $\|\bm{\tilde{\Theta}}\|_2$.
  
 We can write the dual problem of (\ref{DC structure}) as following based on Fenchel's duality \cite{rockafellar2015convex}
  \begin{eqnarray}\label{dual DC structure}
  \mathop{\textrm{minimize}}_{ \bm{Y}} &&q^*(\bm{Y}) - p^*(\bm{Y}),
  \end{eqnarray} 
  where $p^*$ and $q^*$ are the conjugate functions of $p$ and $q$ respectively and matrix $\bm{Y}$ denotes the dual variable. The conjugate function of $f$ is defined as
  \begin{eqnarray}
  p^*(\bm{Y}) = \mathop{\textrm{sup}}_{\bm{Y} \in \mathbb{C}^{m \times n}} \langle \bm{\tilde{\Theta}}, \bm{Y}\rangle -p(\bm{\tilde{\Theta}}),
  \end{eqnarray}
  where $\langle \bm{\tilde{\Theta}}, \bm{Y}\rangle = \mathfrak{R}(\text{Tr}(\bm{\tilde{\Theta}}^\mathsf{H}\bm{Y}))$ and $\mathfrak{R}(\cdot)$ denotes the real part of a complex number \cite{Shi_TSP18demixing}.
  
  Noted that the primal and dual problems are still non-convex. To deal with this challenge, we can exploit successive convex approximation to iteratively update primal and dual variables \cite{Tao1997Convex}. The convex relaxations of the primal and dual problems at the $t$-th iteration are given by
  \begin{align}\label{t-iteration}
  \bm{Y}^t= &\mathop{\textrm{arg\,inf}}_{\bm{Y}}\  q^*(\bm{Y})\!-\!\big[p^*(\bm{Y}^{t-1})\!+\!\langle \bm{Y}\!\!-\!\bm{Y}^{t-1}, \bm{\tilde{\Theta}}^t\rangle\big]\\
  \bm{\tilde{\Theta}}^{t+1} =& \mathop{\textrm{arg\,inf}}_{ \bm{\tilde{\Theta}}}\  p(\bm{\tilde{\Theta}})-\big[q(\bm{\tilde{\Theta}}^{t}) + \langle \bm{\tilde{\Theta}}-\bm{\tilde{\Theta}}^{t}, \bm{Y}^t\rangle\big].
  \end{align}
   
  According to Fenchel biconjugation theorem \cite{rockafellar2015convex}, the above equation (\ref{t-iteration}) can be represented as
  \vspace{-0.10in}
  \begin{align}\label{subgradient}
  &\bm{Y}^t  \in \partial_{\bm{\tilde{\Theta}}^t}q,
  \end{align}
  where $\partial_{\bm{\tilde{\Theta}}^t}q$ is the sub-gradient of $q$ with respect to $\bm{\tilde{\Theta}}$ at $\bm{\tilde{\Theta}}^t$. 
  
  Hence, we have access to $\bm{\tilde{\Theta}}^t$ at the $t$-th iteration via solving the following convex optimization problem:
  \begin{eqnarray}\label{subliftvDC}
  \mathop{\text{minimize}}_{\bm {\tilde{\Theta}}}
\!\!&&\!\text{Tr}(\bm {\tilde{\Theta}})-\langle \bm{{\tilde{\Theta}}},\partial_{\bm{{\tilde{\Theta}}}^{t-1}}\|\bm{{\tilde{\Theta}}}\|_2\rangle\nonumber \\
  \text{subject to}\!\!&&\!\gamma_k \big(\!\sum_{i\neq k}^{K}\text{Tr}(\!\bm{R}_i\bm{\tilde{\Theta}}\!)\!+\!b_{i}^2\!+\!\sigma^2\big)\!\leq\!\text{Tr}(\!\bm{R}_k\bm{\tilde{\Theta}}\!)\!+\!b_{k}^2,  \forall \, k\!\in\! \mathcal{K},\nonumber\\
  &&\sum_{k=1}^{K}c_{l,k}^2+\text{Tr}(\bm{Q}_{l,k}\bm{\tilde{\Theta}})\leq \kappa_l,\forall l\in\mathcal{L},\nonumber\\
  &&\bm{\tilde{\Theta}}_{m,m} =1 , \forall \, m = 1,\ldots,M+1,\nonumber\\
  &&\bm{\tilde{\Theta}} \succcurlyeq 0.
  \end{eqnarray}
  We can adopt CVX to efficiently solve convex problem $\eqref{subliftvDC}$. Specially note that the sub-gradient of  $\|\bm \Psi\|_2$ at $ \bm{\Psi}^{t}$ (i.e., $\bm\partial_{\bm{\Psi}^{t}}\|\bm{\Psi}\|_2$) can be efficiently computed as the method in Proposition 2. 
  \begin{proposition}
        For a PSD matrix $\bm{\Psi} \in \mathbb{C}^{N \times N}$, the sub-gradient of $\|\bm{\Psi}\|_2$ can be efficiently computed as
        \begin{eqnarray}
        \bm{\phi}_1 \bm{\phi}_1^{\sf{H}} \in \partial_{\bm{\Psi}^t}\|\bm{\Psi}\|_2,
        \end{eqnarray}
        where the $\bm{\phi}_1 \in \mathbb{C}^N$ is the eigenvector corresponding to the largest eigenvalue $\lambda_1(\bm{\Psi})$ \cite{Watson1992Characterization}.
  \end{proposition}
  We summarize the presented alternating DC algorithm to solve the original problem $\mathscr{P}$ in Algorithm 1, where problems (\ref{w}) and (\ref{theta}) are solved in an alternating manner until the convergence is achieved. Note that the resulting DC programming problems can be iteratively solved by solving the convex relaxation version of the primal and dual problems of DC programming. We provide simulation results in section $\ref{simu}$ to validate that the proposed  DC approach  outperforms the existing convex methods in terms of transmit power minimization and noise robustness, as the proposed alternating DC approach is far superior, which can guarantee the feasibility of the rank-one constraint.
  
  \begin{algorithm}[htb]
        \SetKwData{Left}{left}\SetKwData{This}{this}\SetKwData{Up}{up}
        \SetKwInOut{Input}{Input}\SetKwInOut{Output}{output}
        \Input{ Initialize $\bm \Theta^1$ and threshold $\epsilon>0$.}
        \For{$t1=1,2,\ldots$}{
                Solve problem (\ref{w}) to obtain  $\bm W^{t1+1}$.\\
                \For{$t =1,2,\ldots$}{
                        Select a subgradient of $\partial \|  \bm{\tilde{\Theta}}^{t-1}\|_2$. \\
                        Solve  problem (\ref{Theta1})  and obtain solution $\bm {\tilde{\Theta}}^t$.\\
                        \If{ \text{Tr}$(\bm {\tilde{\Theta^t}})-\|\bm {\tilde{\Theta^t}}\|_2=$  0}{\textbf{break}}} 
                \If{ the decrease of the total transmit power is below $\epsilon$ or problem (\ref{theta1}) becomes infeasible}{\textbf{break}}}
        \caption{Proposed DC Algorithm for Problem $\mathscr{P}$.}
        \label{algo2}
  \end{algorithm}
 
  \section{Simulation Results}\label{simu}
  In this section, we simulate the proposed alternating DC algorithm in an RIS-assisted cognitive radio network with $K=4$ single-antenna SUs, $L=2$ single-antenna PUs and a 5-antenna SU transmitter. To be specific, the SU transmitter and the RIS are located at $(0,0,10)$ meters and $(50,50,15)$ meters, respectively. The SUs and PUs are randomly distributed in the region of $(-50,50,0)\times (60,160,0)$ and $(-120,-170, 0)\times(-40,10,0)$ meters, respectively. Moreover, we assume the path loss model as $\zeta(d)=T_0(d/d_0)^{-\alpha}$, where $T_0=-30$dB is the path loss with respect to the distance $d_0=1$ meter, $d$ denotes the link distance and $\kappa$ represents the path loss exponent. Specifically, we set the path loss for the Tx-SU link, Tx-PU link, Tx-RIS link, RIS-SU link and RIS-PU link are repetitively set to 3.5, 3.5, 2, 2.2 and 2.2. We further assume Rayleigh fading for all channels such that $\bm{h}=\sqrt{\zeta(d)}\tau$, where $\tau\sim \mathcal{CN}(0,\bm{I})$. In addition, we set $\kappa_l=-30$dB, $\forall  l\in\mathcal{L}$ and $\sigma^2=-80$dB. We then simulate different alternating algorithms denoted as \textit{alternating SDR} and \textit{alternating DC}.
  
  We first study the impart of various values of SINR threshold on the total transmit power. Intuitively, the total transmit power increases with the SINR increasing. It means that we need more transmit power to guarantee the quality-of-service (QoS) for high SINR requirements. In addition, RIS-assisted CR network outperforms traditional CR network without RIS, which further demonstrates the necessity and effectiveness of deploying RIS. Moreover, our proposed alternating DC algorithm achieves smaller transmit power than the alternating SDR.
  
\begin{figure}[t]
        \centering
        \includegraphics[scale = 0.44]{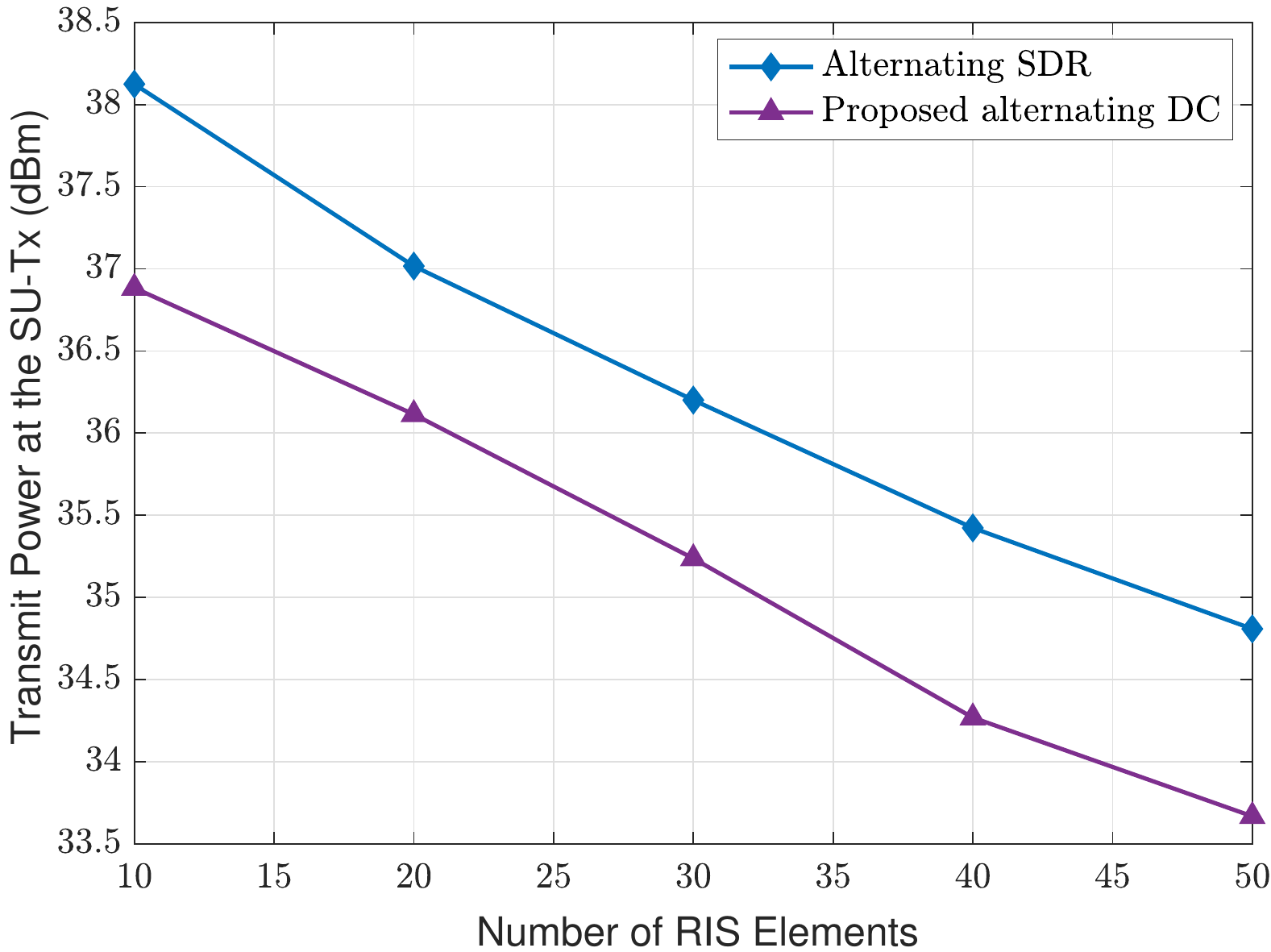}
        \vspace{-0.10in}
        \caption{SU-Tx transmit power versus the number of RIS elements.}
        \label{exp1}
        \vspace{-0.15in}
  \end{figure}

  Fig. \ref{exp1} shows the effectiveness of the number of RIS elements on the total transmit power under the setting $\gamma_k=15$ dB. In general, the total transmit power decreases as the number of RIS elements increases, which indicates that more elements achieve better performance. Moreover, it is clear that our proposed alternating DC algorithm outperforms alternating SDR method.
  \begin{figure}[t]
        \centering
        \includegraphics[scale = 0.44]{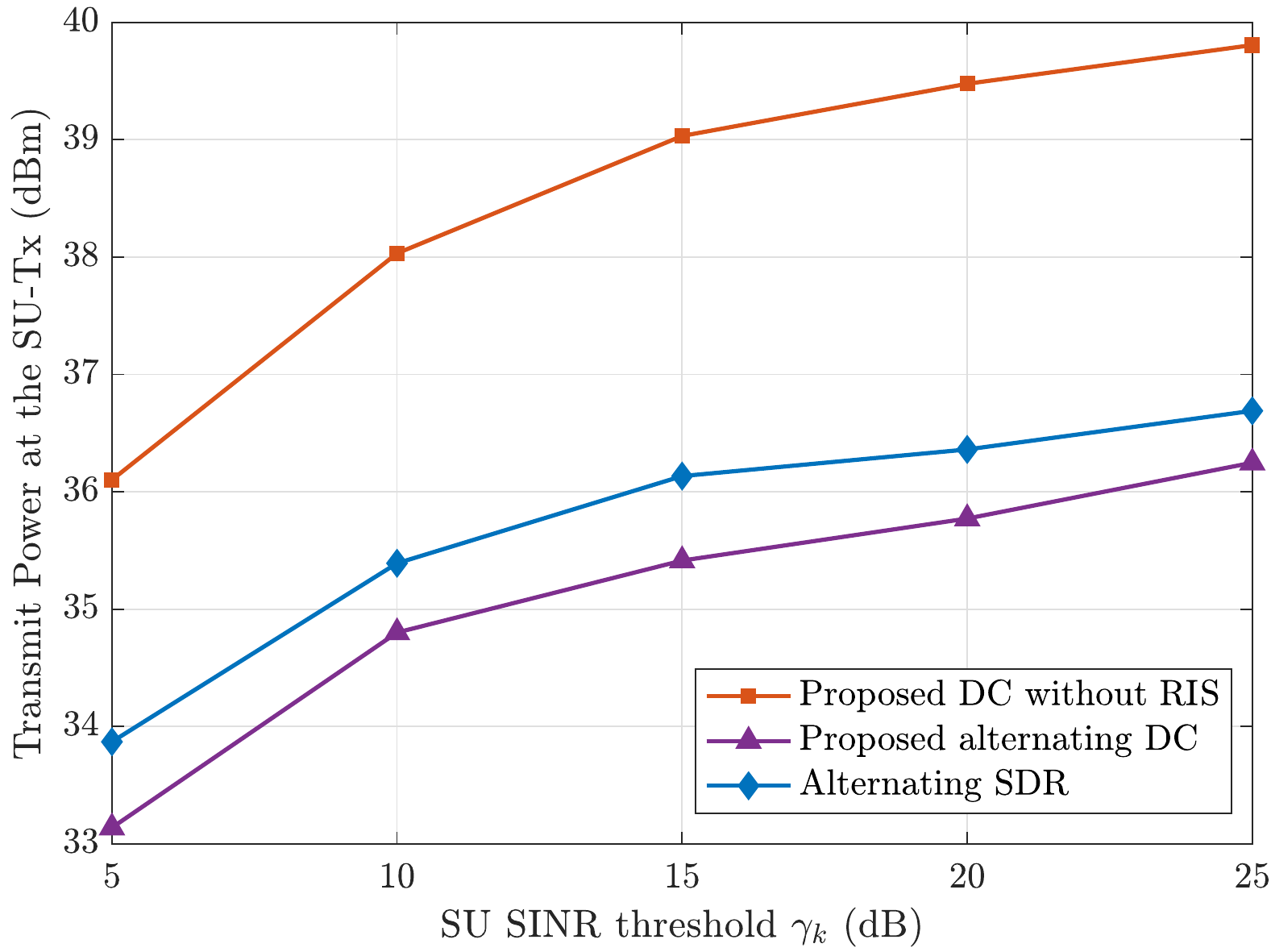}
        \vspace{-0.10in}
        \caption{SU-Tx transmit power versus the SINR target for SUs.}
        \label{exp2}
        \vspace{-0.15in}
  \end{figure}
  
  \section{Conclusions}
  In this paper, we investigated the transmit power minimization problem for the RIS-enhanced CR network by jointly designing the beamforming vectors at the SUs transmitter and the phase shift matrix at the RIS. To this end, we proposed an alternating DC framework to jointly optimize the beamforming vectors and the phase shift matrix in the original problem, and alternatively solved the SOCP optimization subproblem for beamforming vectors and non-convex QCQP subproblem for phase shift matrix at RIS. We then exploited the matrix lifting to reformulate the the non-convex QCQP problem into SDP problem by using an exact DC representation for rank-one constraint. The  DC algorithm was further developed to solve the resulted DC programming problem. Simulation results validated that the performance gains of the proposed alternating DC method versus the existing SDR method.
  
\bibliographystyle{IEEEtran}
\bibliography{IEEEabrv,Reference}
\end{document}